# Realization of a Kondo Insulator in a Multilayer Moiré Superlattice

Qiran Wu[1,#], Jingyuan Cui[1,#], Ang-Kun Wu[2,3,#], Yuze Meng[4,#], Dongxue Chen[5], Li Yan[4,5], Lei Ma[4,5], Takashi Taniguchi[6], Kenji Watanabe[7], Shi-Zeng Lin[2,3,*], Su-Fei Shi[4,*], Yong-Tao Cui[1,*]

1. Department of Physics and Astronomy, University of California at Riverside; Riverside, CA 92521, USA
2. Theoretical Division, T-4, Los Alamos National Laboratory; Los Alamos, New Mexico 87545, USA
3. Center for Integrated Nanotechnologies (CINT), Los Alamos National Laboratory; Los Alamos, New Mexico 87545, USA
4. Department of Physics, Carnegie Mellon University; Pittsburgh, PA 15213, USA
5. Department of Chemical and Biological Engineering, Rensselaer Polytechnic Institute; Troy, NY 12180, USA
6. Research Center for Materials Nanoarchitectonics, National Institute for Materials Science; 1-1 Namiki, Tsukuba 305-0044, Japan
7. Research Center for Electronic and Optical Materials, National Institute for Materials Science; 1-1 Namiki, Tsukuba 305-0044, Japan

[#] These authors contributed equally to this work
[*] Corresponding authors: Email: szl@lanl.gov, sufeis@andrew.cmu.edu, yongtao.cui@ucr.edu

**Kondo insulators are a paradigmatic strongly correlated electron system, arising from the hybridization between itinerary conduction electrons and localized magnetic moments, which opens a gap in the band of conduction electrons. Traditionally, the known Kondo insulators are found in materials with f-electrons. Recent developments in two-dimensional (2D) moiré systems provide a new approach to generate flat bands with strong electron correlation, which host localized moments at half filling. In this work, we demonstrate the realization of a Kondo insulator phase in a moiré superlattice of monolayer $WS_2$ / bilayer $WSe_2$ which hosts a set of moiré flat bands in the $WSe_2$ layer interfacing the $WS_2$ layer and dispersive bands in the other $WSe_2$ layer. When both $WSe_2$ layers are partially doped but with a total density of two holes per moiré unit cell, an insulating state appears when the density of the moiré band is below one hole per moiré unit cell. The insulating state disappears above a certain threshold magnetic field and the system becomes metallic, which is a telltale signature of the Kondo insulator. The physics can be well explained by a periodic Anderson lattice model that includes both the on-site Coulomb repulsion in the moiré flat band and the hybridization between moiré flat and non-moiré dispersive bands. Our results suggest that multilayer moiré structures of transition metal dichalcogenides provide a tunable platform to simulate the Kondo insulator, which holds promise to tackle many critical open questions in the Kondo insulators.**

In a weakly interacting periodic lattice, an electrical insulator forms when the Bloch bands are completely filled by charge carriers, known as a band insulator. When electron interactions are strong, correlated insulators can form at partial band fillings, a famous example being the Mott insulator[1]. The Kondo insulator is another paradigm in strongly correlated electron physics, in which the exchange interaction between localized and itinerant spins opens up a band gap in the Fermi sea[2–7]. It originates from the paradigmatic Kondo effect, which describes the screening of localized magnetic moments by conduction electron spins such that the magnetic moments become part of the itinerant Fermi surface[8]. Extending this Kondo coupling from a single magnetic impurity to a Kondo lattice which hosts a magnetic moment in every unit cell gives rise to the remarkable phenomena: a heavy correlated insulator below the characteristic Kondo temperature when the number of the conduction electrons plus the number of magnetic moments equals two in a unit cell[3,9–11]. Due to the flatness of the band associated with the heavy effective mass, the Kondo lattice is an important platform to explore strongly correlated physics, such as unconventional superconductivity[12–15]. The Kondo insulators have attracted considerable interest recently, largely due to experimental observations of the intriguing quantum oscillations in $SmB_6$[16–18] and $YbB_{12}$[19,20] and their possibilities of being topological Kondo insulators[2,4,21–24]. Progress in understanding these phenomena, however, has been impeded in part by the complexity of these f-electron materials, including their complex band structures and unavoidable impurities. To make progress, it is highly desirable to realize a Kondo insulator simulator with a simple band structure and large tunability.

Recently, 2D moiré systems based on graphene and transition metal dichalcogenides (TMD) have been demonstrated to provide a promising platform to realize various correlated electronic phases, such as superconductivity[25–29], Mott insulators[30–32], generalized Wigner crystals[32–36], ferromagnetism[37–41], integer and fractional quantum anomalous Hall effect[42–49], etc. A key ingredient is the formation of flat bands with strong correlation in the moiré lattice. This provides an ideal playground to simulate the Kondo physics[9,50]. A recent work has demonstrated the formation of heavy fermions in the $MoTe_2/WSe_2$ moiré structure[51]. The $MoTe_2$ layer is doped to the Mott insulator state with one hole per moiré unit cell to act as a lattice of localized moment, and itinerant holes doped in the $WSe_2$ layer are found to behave as conduction electrons. The hybridization of the localized moments and conduction electrons stabilizes a heavy fermion liquid in a range of filling fractions[51]. However, the realization of the Kondo insulator state has remained elusive.

**Design of the Moiré Device Structure for a Kondo Insulator**

In this work, we realize the Kondo insulator by designing a multilayer moiré superlattice structure to couple the moiré flat band (the equivalent of f-electrons) to dispersive bands in a separate layer without any moiré modulation. The structure of our choice is monolayer (1L) $WS_2$ aligned on a natural bilayer (2L) $WSe_2$ (Fig. 1a). Previous research on this system has established the following knowledge about this moiré system. 1) It has a type-II band alignment with the valence bands of $WSe_2$ positioned closer to the Fermi level and well separated from those of $WS_2$[52,53]. 2) The moiré potential is highly localized at the interface between WS$_2$ and WSe$_2$, resulting in moiré flat bands only in the first WSe$_2$ layer at the interface while the other WSe$_2$ layer is not affected[54,55]. 3) The two WSe$_2$ layers in the natural bilayer WSe$_2$ are H-stacked (180 deg). As a result of strong spin-orbit coupling, interlayer tunneling is greatly suppressed between these two layers[56–58]. Therefore, we can consider the electronic structure of valence bands in the bilayer WSe$_2$ as a set of moiré flat bands in the first WSe$_2$ layer and a set of dispersive bands in the second WSe$_2$ layer. An out-of-plane electric field can be used to tune the relative energy alignment between the two WSe$_2$ layers. We take advantage of this feature to control individual doping levels of the two WSe$_2$ layers. Specifically, the first moiré miniband of the first WSe$_2$ layer is tuned to half filling, i.e., one hole per moiré unit cell ($\nu_1 = 1$), which forms an antiferromagnetic Mott insulator[9,31,39,50,51,59,60]. This configuration serves as localized spins to interact with the dispersive band in the second WSe$_2$ layer through spin exchange and forms a Kondo lattice. As a result, when the second WSe$_2$ layer is hole-doped to the same level ($\nu_2 = 1$), the Fermi level is positioned in the Kondo hybridization gap, realizing the Kondo insulator state. We find that this insulating state remains robust in the doping regime of $\nu_1 \leq 1$ and $\nu_2 \geq 1$ as long as $\nu_1 + \nu_2 = 2$ as the consequence of the coherence between the localized and itinerant electrons. This insulating state disappears at a magnetic field above 6.3 T at 6 K, indicating a singlet nature for the hybridized state, the telltale signature of a Kondo insulator.

**Probing the Formation of a Kondo Insulator**

In our experiment, we examine the formation of the insulating state by performing microwave impedance microscopy (MIM) in 1L WS$_2$/2L WSe$_2$ devices with both single and dual gates (see Fig. S1 for optical images of these devices). Fig. 1a is a schematic of a dual-gate device structure. The top gate is a monolayer graphene (MLG), while the bottom gate is a thin graphite flake. MIM measurement is carried out by positioning a sharp metal tip at a fixed spot in the sample region [61–63]. The imaginary and real parts of the complex tip-sample impedance are extracted as the MIM-Im and MIM-Re signals from the MIM electronics (see Methods for details on the MIM measurements). The typical response of MIM-Im and MIM-Re as a function of sample conductivity is plotted in Fig. 1b. In general, the MIM signals have a finite sensitivity window on the sample conductivity, outside of which both channels become saturated. Within the window, the MIM-Im channel changes monotonically with the sample conductivity, which can be used to identify the formation of insulating states. The MIM-Re channel exhibits peaks at an intermediate conductivity with a better sensitivity than MIM-Im when the sample's conductivity is near the saturation values. To enable MIM measurements on dual-gate devices, an out-of-plane magnetic field (2 T and above) is applied to drive the MLG top gate into the quantum Hall regime, which suppresses its bulk conductivity, allowing the microwave electric field from the tip to penetrate through MLG and reach the moiré sample. The MIM signals then correspond to an effective conductivity including both MLG and sample, and we expect to see features from both MLG and sample. Fig. 1d illustrates the expected MIM features from the dual-gated 1L WS$_2$/2L WSe$_2$ device. A set of insulating features at approximately constant voltage on the MLG gate correspond to the

Landau level (LL) gaps in MLG. The insulating features in the TMD moiré device will manifest as diagonal lines in the map, as the carrier density is controlled by both gate voltages. In Fig. 1c, we plot a dual-gate map of MIM-Re at temperature T = 7.8 K and magnetic field B = 3 T. Consistent with our expectations, it exhibits features corresponding to both MLG LLs and insulating states in the TMD moiré device. We identify a few features at several total hole filling values $\nu_{total} = \nu_1 + \nu_2$ as the following: 1) $\nu_{total} = 0$ corresponds to the onset of the hole doping near the charge neutrality gap. 2) At $\nu_{total} = 1$, an insulating state persists through the entire out-of-plane electric field range. Our previous study has demonstrated that this state has a transition from a Mott insulator to an excitonic insulator[54]. When the electric field pushes all the holes to the first WSe$_2$ layer (the upper portion of the $v_{total} = 1$ line in Fig. 1c), it corresponds to a Mott insulator. When some of the holes are pushed to the second WSe$_2$ layer (the lower portion of the $v_{total} = 1$ line in Fig. 1c), holes in the first and second WSe$_2$ layers are correlated to form an excitonic insulator with $v_1 + v_2 = 1$ [54,64,65]. 3) At $v_{total} = 2$, insulating states appear at two opposite ends where the direction of the electric field is opposite. At the upper-left branch, the electric field pushes all the holes into the first WSe$_2$ layer $v_1 = 2$, leaving the second WSe$_2$ layer empty $v_2 = 0$, and it corresponds to the band insulator state where the first moiré hole band in the first WSe$_2$ layer is completely filled (Fig. 1e). The point A in Fig. 1c marks the onset of this state, $(v_1, v_2) = (2,0)$. By tuning only the top gate, we go horizontally in Fig. 1c from point A to point B on the $v_{total} = 1$ line, and $v_2$ should remain 0 with a constant bottom gate voltage. The point B thus corresponds to $(v_1, v_2) = (1,0)$, which is a Mott insulator (Fig. 1f). Now further going vertically in Fig. 1c from B to C, which is the onset of the other insulating state on the $v_{total} = 2$ line, we can see that B and C have approximately the same top gate voltage thus the same $v_1 = 1$, and so the point C is the onset of $(v_1, v_2) = (1,1)$, i.e., the first WSe$_2$ layer is at the Mott insulator state while the second WSe$_2$ layer hosts the same number of holes (Fig. 1g). According to our previous analysis, this is the condition to form a Kondo insulator, and the MIM data clearly shows an insulating state starting at C. The insulating state persists over a range of the electric field which could extend to the regime of $v_1 < 1, v_2 > 1$ while maintaining $v_{total} = v_1 + v_2 = 2$.

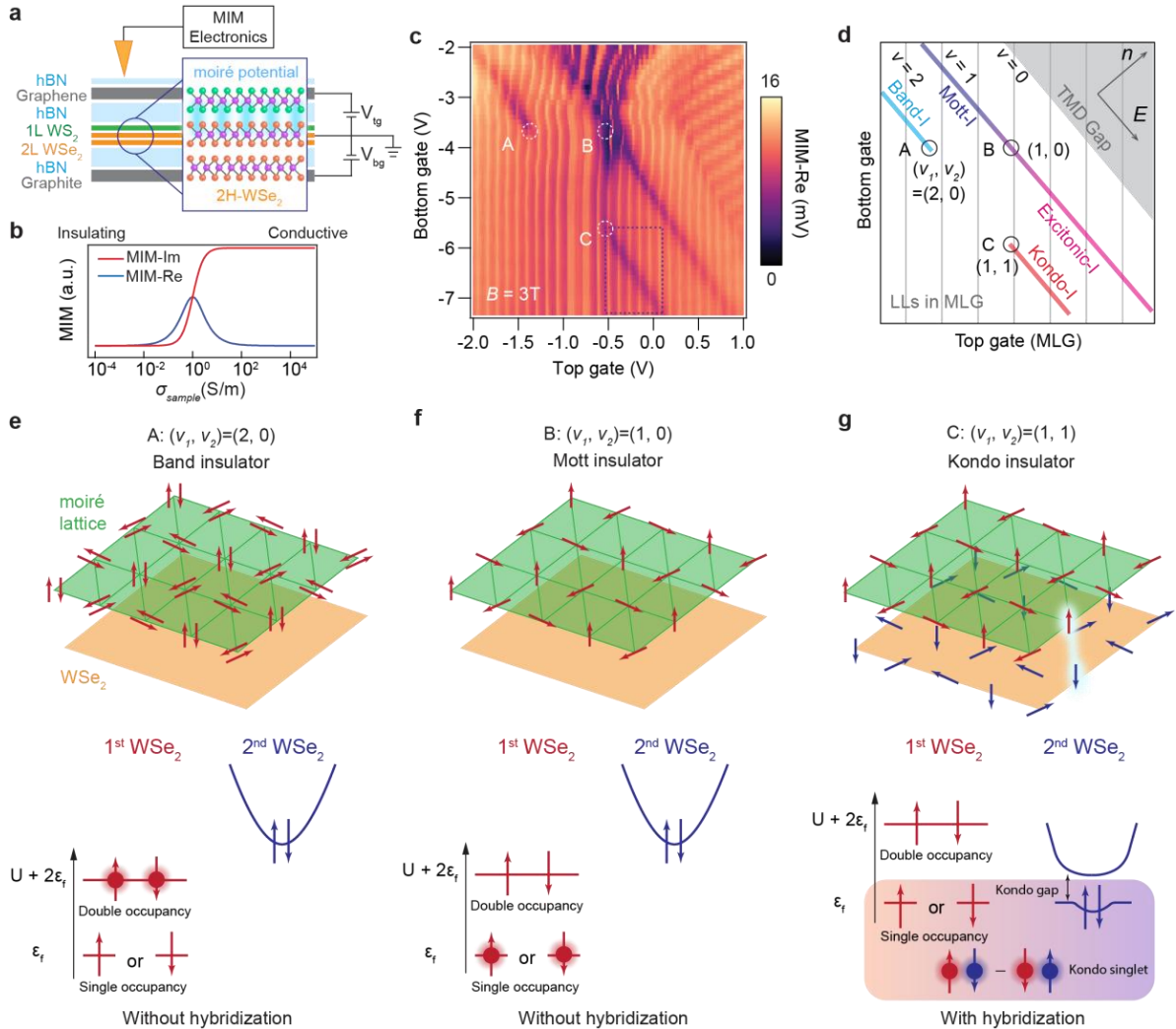

**Figure 1. Kondo insulator in the 1L WS$_2$/2L WSe$_2$ moiré superlattice structure. a,** Schematic of the 1L WS$_2$/2L WSe$_2$ device structure with dual gates. The moiré potential is highly confined at the interface between WS$_2$ and WSe$_2$. **b,** Typical response curves of MIM-Im and MIM-Re as a function of samples' conductivity. **c,** Dual-gate map of MIM-Re at a single spot on 1L WS$_2$/2L WSe$_2$ at B = 3 T and T = 7.8 K. **d,** Schematic of the major features in a dual-gate map. The nature of the insulating states is labeled at different density configurations. **e-g,** Illustrations of carrier configurations (top) and band diagrams (bottom) at points (**e**) A, (**f**) B, and (**g**) C, as marked in **c** and **d**.

**The Collapse of the Kondo Insulator at High Magnetic Field**

In the Kondo insulator, the hybridization between the localized electrons in the moiré layer and the conduction electrons leads to the formation of the Kondo singlet. A large magnetic field can break the Kondo singlet when the Zeeman energy exceeds the Kondo insulator gap. As a consequence, the Kondo hybridization gap collapses, and the system transitions from the Kondo insulator to a metal. To examine this gap closing, we measure the dual gate map of MIM signals at B=9 T (Fig. 2a). The insulating state in the $v_1 \leq 1, v_2 \geq 1$ region no longer exists, while the band insulator at $v_{total} = 2$ and the Mott/excitonic insulator at $v_{total} = 1$ remain. A detailed magnetic field dependence indicates that the Kondo insulator state breaks down at around B=6 T (see Fig. S2). To study the magnetic field evolution in more detail, we turn to single gate devices without any top gates, which enhances the MIM signals. The MIM vs bottom gate voltage trace is equivalent to a line cut along zero top gate voltage which indeed goes through the Kondo insulator state. Fig. 2b plot the MIM traces taken at zero field, and the insulating feature at $v_{total} = 2$ is clearly present in both MIM-Im and MIM-Re traces, along with the excitonic insulator state at $v_{total} = 1$, and the generalized Wigner crystal states at fractional fillings of $v_{total} = 1/3$ and $2/3$.

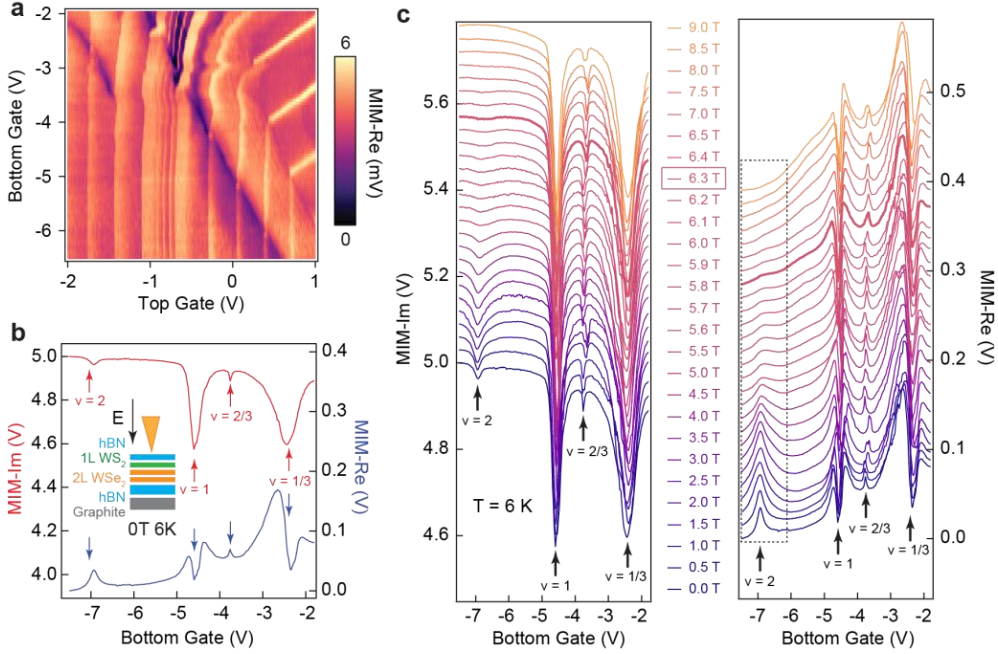

**Figure 2. Breakdown of the Kondo insulator phase by magnetic field. a,** Dual-gate MIM-Re map taken at the same spot as in Fig. 1c at B = 9 T and T = 6.8 K. **b,** MIM-Im and MIM-Re as a function of bottom gate voltage in a single-gate 1L WS$_2$/2L WSe$_2$ device at B = 0 T and T = 6 K. The inset shows the device structure. **c,** The MIM-Im and MIM-Re as a function of bottom gate voltage taken at selected magnetic fields as labelled from 0 T to 9 T. The insulating feature at $v_{total} = 2$ disappears at B = 6.3 T at a temperature of 6 K, while other correlated insulating states remain unchanged. The critical field is determined using MIM-Re data within the dashed-line region. Both MIM channels are vertically offset for clarity.

Fig. 2c plots the MIM-Im and MIM-Re gate sweeps at magnetic fields from 0 to 9 T. The Kondo insulator feature disappears at B = 6.3 T (see Fig. S3), while other correlated insulating states barely have any changes. The corresponding Zeeman energy is estimated to be 3.6 meV (with a g-factor of 10 for WSe$_2$[66]), providing an estimate of the Kondo insulator gap.

**Transition Temperature of the Kondo Insulator**

We further characterize the temperature dependence of the Kondo insulator states. Fig. 3a plots the MIM-Im and MIM-Re vs bottom gate voltage in the single gate device at temperatures from 6 K to 40 K at zero magnetic field. The Kondo insulator feature at $v_{total} = 2$ disappears at around T$_c$ = 35 K (see Fig. S3). Above this temperature, the hybridization between the localized moments and conduction electrons loses coherence, and the insulating gap disappears. The value of $k_B T_c =$ 3 meV provides another estimate of the Kondo insulator gap at zero field. Fig. 3b plots the transition temperatures at different magnetic fields (see the complete data set in Fig. S4 and Fig. S5). T$_c$ drops quickly above 5 T, consistent with the quickly diminishing MIM signals in Fig. 2c.

## Measurement of the Kondo Gap

To directly extract the energy gap at the Kondo insulator state, we utilize the MLG top gate in the dual-gate device as a sensor for probing the chemical potential in the moiré device. Fig. 3c plots a high-resolution MIM-Re map around the Kondo insulator region where multiple LLs cross the Kondo insulator state. By tracking the gate voltage position of individual LLs, we can determine the chemical potential of the TMD moiré device as a function of the carrier density, using a methodology we have demonstrated previously[67]. The four traces in Fig. 3d are extracted from four LL's which intersect the $\nu_{total} = 2$ line at different electric fields and overlaid together. The jump in chemical potential at $\nu_{total} = 2$ in each curve corresponds to the thermodynamic gap of the Kondo insulator, which is around 3-4 meV, consistent with our estimates based on the critical magnetic field at 6 K and transition temperature at 0 T. The overall negative slope suggests a negative electronic compressibility, as demonstrated in our previous experiment on the 1L WS$_2$ / 1L WSe$_2$ moiré structure[67], which is also typical in other systems with strong electron interactions[68,69]. The four curves at different electric fields overlap nicely along a similar profile, indicating that the Kondo insulator gap has no obvious dependence on the electric field in this range. Fig. S6 plots measurements at different spots in this device which show a variation of the gap value between 3-7 meV.

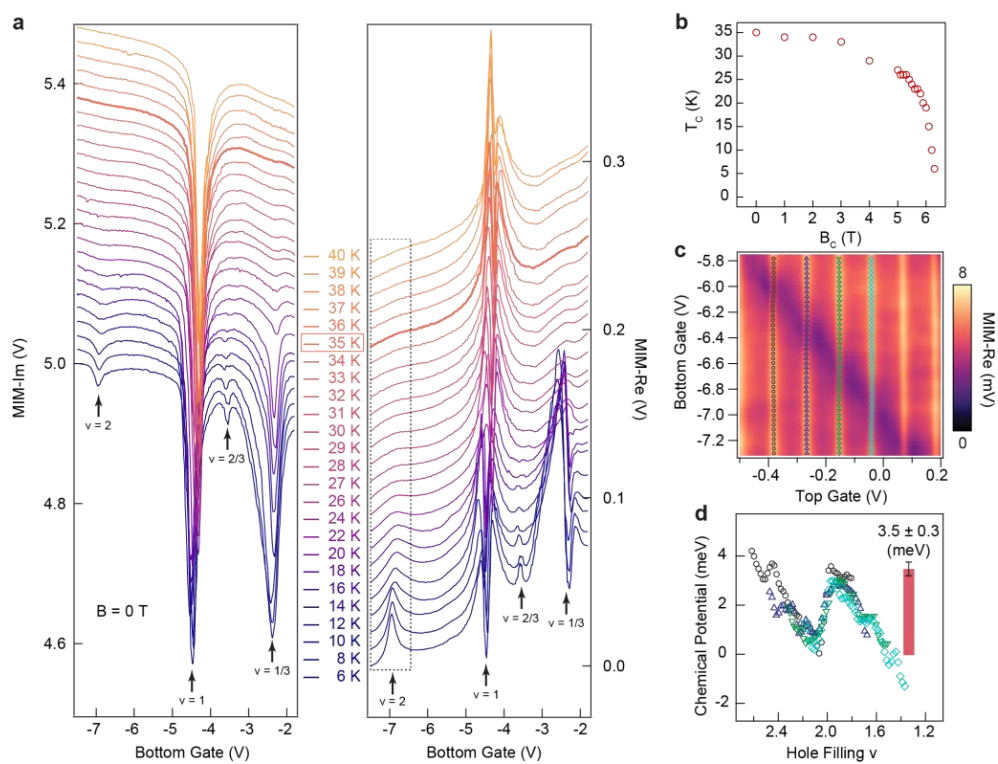

**Figure 3. Transition temperature and thermodynamic energy gap of the Kondo insulator phase. a,** MIM-Im and MIM-Re as a function of bottom gate voltage at temperatures from 6 K to 30 K at zero magnetic field. The insulating feature at $\nu_{total} = 2$ disappears at a transition temperature around 35 K. The transition temperature is determined using MIM-Re data within the dashed-line region. Both MIM channels are vertically offset for clarity. **b,** Transition temperature as a function of the magnetic field. **c,** High resolution dual-gate MIM-Re map taken in the same device as Fig. 1c in a fine voltage range around the Kondo insulator phase. The marked lines trace the position of the Landau levels in the MLG top gate. **d,** The chemical potential of the moiré device extracted from the Landau levels positions in (**c**) plotted as a function of hole density. The curves are matched to have the same chemical potential at $\nu_{total} = 2$.

**Theoretical modeling of the Kondo Insulator**

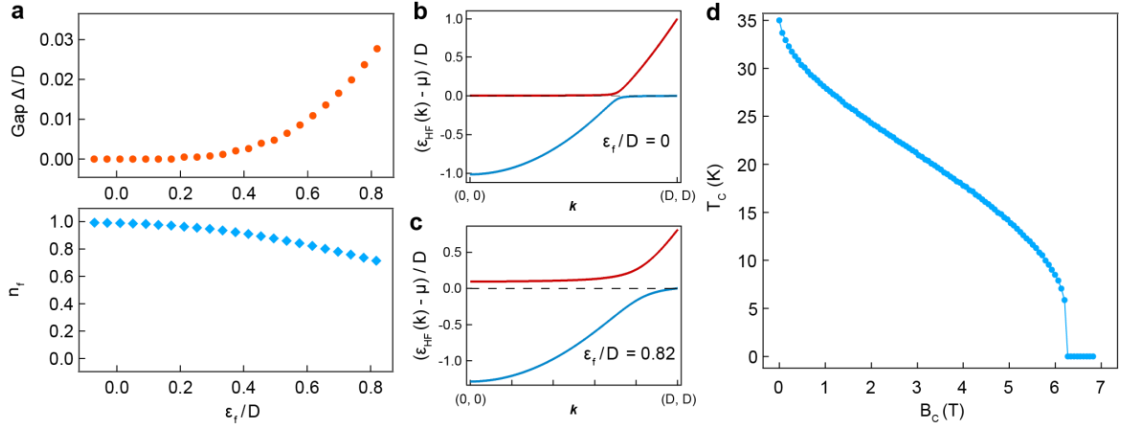

**Figure 4. Theoretical results from slave-boson mean-field calculations. a,** The calculated Kondo Insulator gap $\Delta$ and local moment density $n_f$ as a function of local moment energy $\varepsilon_f$ relative to the bottom of the dispersive band. The energy is normalized to a bandwidth $D$ corresponding to filling the dispersive band with carriers at the moiré density, $\nu_2 = 1$. The total particle filling is fixed at $\nu_{\text{total}} = 2$ per site by tuning the chemical potential $\mu$. **b,c,** Dispersions of the hybridized bands at (**b**) $\varepsilon_f = 0$ and (**c**) 0.82$D$. **d,** The mean-field critical temperature (when $\Delta \to 0$) as a function of critical magnetic fields with fixed chemical potential $\mu = 0$, $\varepsilon_f = 0$. The bare hybridization is $V = 0.4D$. We take $D$=32.8 meV in calculations for (**d**).

To analyze the formation of the Kondo insulator, we treat the 2L WSe$_2$ as a simple two-band system in which each layer contributes one. In the first layer, the flat moiré band can accommodate 0, 1, or 2 carriers per unit cell. The double occupancy upper Hubbard band (UHB) is separated from the single occupancy lower Hubbard band (LHB) by the large on-site Coulomb repulsion, $U$. The second layer hosts a regular dispersive hole band whose energy can be tuned relative to the moiré band. In the presence of a large negative electric field that puts the dispersive band at a much higher energy (for holes) than the moiré band, the carriers are strongly polarized to fully fill the moiré band at $\nu_{\text{total}} = 2$, giving rise to a trivial band insulator (see Fig. 1e corresponding to the insulating region onset at point A in Fig. 1c). As the electric field brings the dispersive band to overlap with the UHB, part of the carriers is transferred to the dispersive band, resulting in a gapless phase (the metallic region along the line between A and C in Fig. 1c). As the dispersive band is further lowered, the lower energy physics only involves the single-occupancy LHB ($\nu_1 = 1$) and the dispersive band, and the interaction can be described by the second-order spin exchange, realizing the paradigmatic Kondo lattice. As a result, their hybridization opens a gap in the dispersive band at $\nu_{total} = 2$, and a Kondo insulator is formed (Fig. 1g). This configuration corresponds to the insulating region started from point C in Fig. 1c. The size of the Kondo insulator gap depends on the antiferromagnetic Kondo coupling strength, $J$. As the electric field further increases, the LHB and the dispersive band become close in energy, allowing charge fluctuations or even partial filling in the LHB. In this case, the hybridization still remains, corresponding to the regime of $\nu_1 \leq 1$ and $\nu_2 \geq 1$.

To corroborate the above picture, we model our device using the periodic Anderson lattice model, which describes the coupling of carriers between the two bands as a spin-independent tunneling term (see Methods for details of the model). The energy of the LHB (the equivalent of f-electron) relative to the dispersive band (c-electron) is tuned by the parameter $\varepsilon_f$, where $\varepsilon_f = 0$ corresponds to the energy position at the bottom of the dispersive band. We use the slave boson approach[70,71] to obtain the band structure. A non-zero slave boson parameter $\langle b \rangle \neq 0$ correponds to a nonzero hybridization between the two bands, which opens a hybridization gap $\Delta$, and the carrier density on the LHB is $n_f = 1 - \langle b \rangle$. The calculated results of $\Delta$ and $n_f$ as a function of $\varepsilon_f$ are plotted in Fig. 4a. As $\varepsilon_f$ increases, $\Delta$ also grows, signaling the formation of the Kondo insulator phase, while $n_f$ drops below 1 as a result of the hybridization. The calculated band dispersions of the hybridized states are plotted in Fig. 4b and 4c for two values of $\varepsilon_f$. In both cases, flat bands form near the Fermi level, resulting in a large density of states near the gap, known as the Kondo resonance[72]. The temperature-magnetic field phase diagram in the Kondo insulator phase is calculated and presented in Fig. 4d, which is qualitatively in agreement with the experimental observation.

To summarize, we have realized the Kondo insulator phase in a multilayer moiré superlattice by separately creating the localized moment and itinerant electrons: the localized moment is provided by the moiré superlattice of $WS_2/WSe_2$ with strong electronic correlation, while the conduction electron resides in a separate $WSe_2$ layer. The key to enabling the Kondo coupling in this device structure is the strongly suppressed interlayer tunneling between the bilayer $WSe_2$, which is made possible by adopting an H-stacked bilayer. This design principle can be applied to similar moiré structures to explore Kondo-related physics. Here we have focused on the gapped phase, and it is natural to expect that our device also realizes the heavy fermion liquid phase when doped away from the Kondo insulator state. Therefore, the multilayer moiré superlattice provides a versatile platform to simulate Kondo lattice physics, which holds promise to resolve many important open questions in Kondo systems.

**Methods**

**Moiré device fabrication.** We use a dry pickup method to fabricate the $WS_2/WSe_2$ heterostructures. We exfoliate monolayer $WS_2$, multilayer $WSe_2$, MLG, few-layer graphite, and hBN flakes on a silicon substrate with a 285 nm thermal oxide layer. For angle-aligned heterostructures, we choose exfoliated WS$_2$ and WSe$_2$ flakes with sharp edges, whose crystal axes are further confirmed by second harmonic generation (SHG) measurements. We then mount the SiO$_2$/Si substrate on a rotational stage and clamp the glass slide with thin flakes to another three-dimensional (3D) stage. We adjust the 3D stage to control the distance between substrates and thin flakes, and we sequentially pick up different flakes onto the pre-patterned Pt electrodes on SiO$_2$/Si substrates. We fine adjust the angle of the rotational stage (accuracy of 0.02°) under a microscope objective to stack the WS$_2$/WSe$_2$ heterojunction, ensuring a near-zero twist angle between the two flakes. The final constructed device is annealed at 130 °C for 12 hours in a vacuum chamber. The pre-patterned Pt contact electrodes are fabricated through standard electron-beam lithography and e-beam evaporation processes (See Fig. S1 for optical microscope images of the devices D1, D2 and D3). In dual-gate devices, monolayer graphene flakes were transferred over the device stack to serve as the top gate.

**Microwave impedance microscopy measurements.** The MIM measurement is performed on a homebuilt cryogenic scanning probe microscope platform. A small microwave excitation of about 0.1 µW at a fixed frequency around 10 GHz is delivered to a chemically etched tungsten tip mounted on a quartz tuning fork. The reflected signal is analyzed to extract the demodulated output channels, MIM-Im and MIM-Re, which are proportional to the imaginary and real parts of the admittance between the tip and sample, respectively. To enhance the MIM signal quality, the tip on the tuning fork is excited to oscillate at a frequency of around 32 kHz with an amplitude of approximately 8 nm. The resulting oscillation amplitudes of MIM-Im and MIM-Re are then extracted using a lock-in amplifier to yield d(MIM-Im)/dz and d(MIM-Re)/dz, respectively. The d(MIM)/dz signals are free of fluctuating backgrounds, and their behavior is very similar to that of the standard MIM signals. In this paper we simply refer to d(MIM)/dz as the MIM signal.

**Determination of critical field and transition temperatures.** The critical field at 6 K and transition temperatures for Kondo insulator phases are extracted from single gated device D2. A spatial MIM scan is performed before each gate voltage sweep to ensure consistent measurement alignment on the same spot, mitigating drift from magnetic field or temperature variations. To minimize ambiguity caused by noise or thermal drift, MIM data are normalized to the MIM-Im signal range (maximum signal – minimum signal), which reflects the signal contrast between the most conducting and insulating states of the sample and should remain constant across measurements. The MIM-Re channel data around $v = 2$ is fitted to a Gaussian function with a linear background: $V_{MIM-Re} = a\frac{(V_B-b)^2}{2c^2} + kV_B + m$, where the normalized peak height $a$ is used to determine the critical parameters. The critical magnetic field or transition temperature is defined as the point where $a$ falls below 0.8% of the MIM-Im signal range, a threshold corresponding to the typical signal noise level observed in the measurements.

**Determination of Kondo gap.** The Kondo gap is determined from the abrupt Landau level (LL) shifts in the MLG top gate as they cross the Kondo insulating phase[67]. The carrier density in MLG ($n_G$) depends on the potential difference between the MLG and the TMD heterotrilayer, given by

$$n_G = C_{tg}\left(\frac{\mu_{TMD} - \mu_G}{e} - V_{tg}\right)$$

where $C_{tg}$ is the geometric capacitance between the top gate and the TMD sample, $\mu_{TMD}$ and $\mu_G$ are the chemical potentials of the sample and MLG, e is electron charge, and $V_{tg}$ is the top-gate voltage. When MLG is kept in a specific LL gap ($n_G$, $\mu_G$ constant), $\Delta\mu_{moire} = e\Delta V_{tg}$. We thus can extract the $V_{tg}$ value for a particular LL as a function of $V_{bg}$, which is then converted to $\Delta\mu_{moire}$ vs $\nu_{total}$.

The positions of Landau levels in dual-gate map of MIM-Re are sharply defined. The Kondo gap is calculated through the following steps: First, we split the $(V_{tg}, V_{bg})$ coordinates along a Landau level into two groups at a sharp jump of $V_{tg}$. The midpoint between groups defines the split point $(V_{tg0}, V_{bg0})$. Second, each group is fitted separately to a linear model $V_{bg} = aV_{tg} + b$, yielding fitting parameters $(a_1, b_1)$ and $(a_2, b_2)$. Third, the gap at the split point $(V_{tg0}, V_{bg0})$ is calculated as $\Delta = |(a_2 V_{tg0} + b_2) - (a_1 V_{tg0} + b_1)|$. The electrical field is given by $E = 1/2(V_{tg0}/d_t - V_{bg0}/d_b)$, where $d_t$ and $d_b$ are the top and bottom hBN thicknesses. Finally, the uncertainties for both fits ($\delta_1$, $\delta_2$) are combined to determine the gap uncertainty $\delta\Delta = \sqrt{\delta_1^2 + \delta_2^2}$.

With no electric field dependence observed, the Kondo gap is averaged across measurements. Using weights $\omega_i = 1/\delta\Delta_i^2$, the weighted average gap size $\bar{\Delta}$ and its uncertainty $\bar{\sigma}$ are:

$$\bar{\Delta} = \frac{\sum_{i=1}^n \omega_i \Delta_i}{\sum_{i=1}^n \omega_i}, \quad \bar{\sigma} = \frac{1}{\sqrt{\sum_{i=1}^n \omega_i}}$$

**Model calculation.** The Kondo lattice physics can be modeled by the periodic Anderson model (PAM)

$$H_{\text{PAM}} = \sum_{k\sigma}(\epsilon_k - \mu)c_{k\sigma}^\dagger c_{k\sigma} + \sum_{j\sigma}(\epsilon_f - \mu)f_{j\sigma}^\dagger f_{j\sigma} + U\sum_j n_{j\uparrow}^f n_{j\downarrow}^f + V\sum_{j\sigma}(f_{j\sigma}^\dagger c_{j\sigma} + c_{j\sigma}^\dagger f_{j\sigma}),$$

where $c_{k\sigma}, c_{j\sigma}, f_{j\sigma}$ are itinerant electron ($c$) and localized electron ($f$, Mott band) annihilation operators at momentum k or site $j$ with spin σ, $\epsilon_k$ the dispersion of the itinerant electrons, $\epsilon_f$ the energy of the localized electrons, and μ the chemical potential. V denotes the hybridization between localized and itinerant electrons at each local site $j$, and $U$ is the onsite repulsive interaction of localized electrons with $n_{j\sigma}^f = f_{j\sigma}^\dagger f_{j\sigma}$. In the large $U$ limit and when the *f*-electron level is singly occupied, the PAM can be written as the Kondo lattice Hamiltonian using the Schrieffer-Wolff transformation[73].

To solve the interacting Hamiltonian, we take the large $U$ limit and use the slave-boson mean-field approach. The slave-boson approach captures the fluctuations of the *f*-electron between single-occupancy and zero-occupancy via a boson field and reduces the many-body problem to a mean-field single-particle problem with self-consistency. The effect of magnetic field is introduced through a Zeeman coupling term, and the effect of temperature is modelled by the Fermi-Dirac distribution. More details on the model calculation can be found in Section 8 of the Supplementary Information.

## Data availability

Source data are available for this paper. All other data that support the plots within this paper and other findings of this study are available from the corresponding author upon reasonable request.

## Acknowledgments

Q.W., J.C., and Y.-T.C. acknowledge support from NSF under award DMR-2145735. S.-F.S. acknowledges the support from NSF (Career Grant DMR-1945420, DMR-2104902 and ECCS-2139692). K.W. and T.T. acknowledge support from the Elemental Strategy Initiative conducted by the MEXT, Japan, Grant Numbers JPMXP0112101001 and JSPS KAKENHI, Grant Numbers 19H05790 and JP20H00354. The work at LANL was carried out under the auspices of the U.S. DOE NNSA under contract No. 89233218CNA000001 through the LDRD Program, and was performed, in part, at the Center for Integrated Nanotechnologies, an Office of Science User Facility operated for the U.S. DOE Office of Science, under user proposals #2018BU0010 and #2018BU0083.

## Author Contributions

Y.-T.C., S.-F.S., and S.-Z.L. conceived and supervised the project. Q.W. and J.C. carried out the MIM measurement and data analysis. Y.M and D.C. fabricated the moiré devices. A.-K.W. performed the numerical calculations. L.Y. and L.M. assisted in the optical characterization of the devices. T.T. and K.W. provided the hBN crystals. Q.W., A.-K.W., S.-Z.L., and Y.-T. C. wrote the manuscript with inputs from all authors.

## Competing Interests

The authors declare no competing interests.